\documentclass[12pt]{article}
\usepackage{epsfig}
\usepackage{latexsym}
\usepackage{amstext}
\usepackage{amssymb}
\usepackage{graphics}

\newcommand{\ds}{\displaystyle}
\newcommand{\beq}{\begin{eqnarray}}
\newcommand{\eeq}{\end{eqnarray}}
\newcommand{\beqq}{\begin{eqnarray*}}
\newcommand{\eeqq}{\end{eqnarray*}}

\begin{document}
\pagestyle{plain} \
\begin{center}{\bf STOCHASTIC RESONANCE OF ELF-EMF IN VOLTAGE-GATED CHANNELS:
THE CASE OF THE CARDIAC $I_{\mbox{\scriptsize Ks}}$ POTASSIUM CHANNEL}\\[3mm]
M. Shaked\footnote{Partially supported by a research grant from
the Adams Brain Research Center, Tel-Aviv University}\\[3mm]

Department of Systems, School of Electrical Engineering, The
Fleischman Faculty of Engineering, Tel-Aviv University, Ramat-Aviv
Tel-Aviv 69978, Israel\\
e-mail: shakedm@eng.tau.ac.il\\[5mm]

G. Gibor\\[3mm]

The Sackler Medical School, Department of Physiology and Pharmacology,
Tel-Aviv University, Ramat-Aviv Tel-Aviv 69978, Israel\\
e-mail: \\[5mm]

B. Attali\\[3mm]

The Sackler Medical School, Department of Physiology and
Pharmacology, Tel-Aviv University, Ramat-Aviv Tel-Aviv 69978, Israel\\
e-mail: battali@post.tau.ac.il\\[5mm]

Z. Schuss\footnote{Partially supported by research grants from the
Israel Science Foundation and the US-Israel Bi-national Science Foundation}\\[3mm]
Department of Mathematics, Tel-Aviv-University, Ramat-Aviv, Tel-Aviv
69978, Israel\\
e-mail: schuss@post.tau.ac.il\\[1cm]

{\bf ABSTRACT}
\end{center}

\noindent We applied a periodic magnetic field of frequency 16 Hz
and amplitude 16 nT to a human $I_{KS}$  channel, expressed in a
{\em Xenopus} oocyte and varied the membrane depolarization between
-100 mV and +100 mV. We observed a maximal increase of about 9\% in
the potassium current at membrane depolarization between 0 mV and 8
mV (see Figure \ref{f:SIKUM_Iks_change_B_16nT_f16_sofi}). A similar
measurement of the potassium current in the KCNQ1 channel, expressed
in an oocyte, gave a maximal increase of 16\% at the same applied
magnetic field and membrane depolarization between -14 mV and -7 mV
(see Figure \ref{f:SIKUM_KCNQ1_change_B_16nT_f16_sofi}). We
attribute this resonant behavior to stochastic resonance between the
thermal activations of the configuration of interacting ions in the
$I_{\mbox{\scriptsize Ks}}$ channel over a low potential barrier
inside the closed state of the channel and the periodic
electromotive force induced across the membrane by the periodic
magnetic field. The partial synchronization of the random jumps with
the periodic force changes the relative times spent on either side
of the barrier, thereby changing the open probability of the
spontaneously gating open channel. This, in turn, changes the
conductance of the channel at the particular depolarization level
and frequency and is expressed in the Hodgkin-Huxley equations as a
bump at the given voltage in the conductance-voltage relation. We
integrate the modified Hodgkin-Huxley equations for the
$I_{\mbox{\scriptsize Ks}}$ current into the Luo-Rudy model of a
Guinea pig ventricular cardiac myocyte and obtain increased
conductance during the plateau of the action potential in the cell.
This shortens both the action potential and the cytosolic calcium
concentration spike durations, lowers its amplitude, increases
cytosolic sodium, and lowers cytosolic potassium concentrations. The
shortening of the ventricular calcium signal shortens the QT period
of the ECG. These theoretical predictions, supported by experimental
measurements, show that $I_{\mbox{\scriptsize Ks}}$-current boosting
by ELF-EMF may be beneficial for increasing the repolarization
reserve and thereby for preventing the prolongation of the action
potential and the risk of ventricular arrhythmias.

\section{Introduction}

The recent spate of communications on ELF-EMF effects on heart
\cite{DiCarlo}, \cite{Sait}, \cite{Fadel}, \cite{Jeong2005}, on
cardiac myocytes in vitro \cite{Asher2007}, on cardiovascular
disease mortality \cite{Savitz}, on human blood pressure
\cite{Korpinen}, on increase in vitro and in vivo of angiogenesis
\cite{Tepper}, on calcium-ion efflux from brain tissue in vitro
\cite{Blackman}, \cite{Blackman85}, on human leukemia T-cells
\cite{Galvanovskis}, on the control of neutrophil metabolism
\cite{Rosenspire}, on cell membranes \cite{Baureus}, on
characteristics of membrane ions \cite{McLeod}, on system of ions
\cite{Giudice}, on ion cyclotron resonance \cite{Liboff}, on ion
thermal motion in a macromolecule \cite{Zhadin}, \cite{Adair},
raises the question of pinpointing the source of the interactions at
the molecular level.

The immediate suspects of the said interaction are the KCNQ1
channels (Kv7.1) that belong to a subfamily of voltage-gated K$^+$
channels, Kv7, and co-assemble with KCNE1 $\beta$ subunits to
generate the $I_{\mbox{\scriptsize Ks}}$ potassium current that is
critical for normal repolarization of the cardiac action potential
\cite{Sanguinetti1991}, \cite{Barhanin1996}, \cite{Barhanin2000},
\cite{Jentsch2000}, \cite{Nerbonne2005}. The reason for suspecting
the $I_{\mbox{\scriptsize Ks}}$ channel of complicity in affecting
calcium transients in cardiac myocytes  and in the general cardiac
response to ELF-EMF, such as  the shortening shortening of the QT
interval (mutations in either KCNQ1 or KCNE1 genes produce the long
QT syndrome (LQT) \cite{Mazhari2002}), a human ventricular
arrhythmia \cite{Jentsch2000}, \cite{Nerbonne2005},
\cite{Roden2006}, and similar phenomena communicated in
\cite{Mazhari2002}, \cite{Jeong2005}, \cite{Asher2007}, is that they
stay open for the duration of the plateau in the action potential of
cardiac myocytes. This leaves sufficient time for the
$I_{\mbox{\scriptsize Ks}}$ current to interact with the ELF-EMF.

The main result of this paper is to pinpoint the interaction of
the ELF-EMF at the suspected KCNQ1 and $I_{\mbox{\scriptsize Ks}}$
channels. To do this, we applied a periodic magnetic field of
frequency 16 Hz and amplitude 16 nT to a human
$I_{\mbox{\scriptsize Ks}}$ channel, expressed in a {\em Xenopus}
oocyte and varied the membrane depolarization between -100 mV and
+100 mV. We observed a maximal increase of about 9\% in the
potassium current at membrane depolarization between 0 mV and 8 mV
(see Figure \ref{f:SIKUM_Iks_change_B_16nT_f16_sofi}). A similar
measurement of the potassium current in the KCNQ1 channel,
expressed in an oocyte, gave a maximal increase of 16\% at the
same applied magnetic field and membrane depolarization between
-14 mV and -7 mV (see Figure
\ref{f:SIKUM_KCNQ1_change_B_16nT_f16_sofi}).

\begin{figure}
\mbox{
\begin{minipage} {\textwidth}
\begin{center}
\begin{tabular}{c}
\epsfig {figure=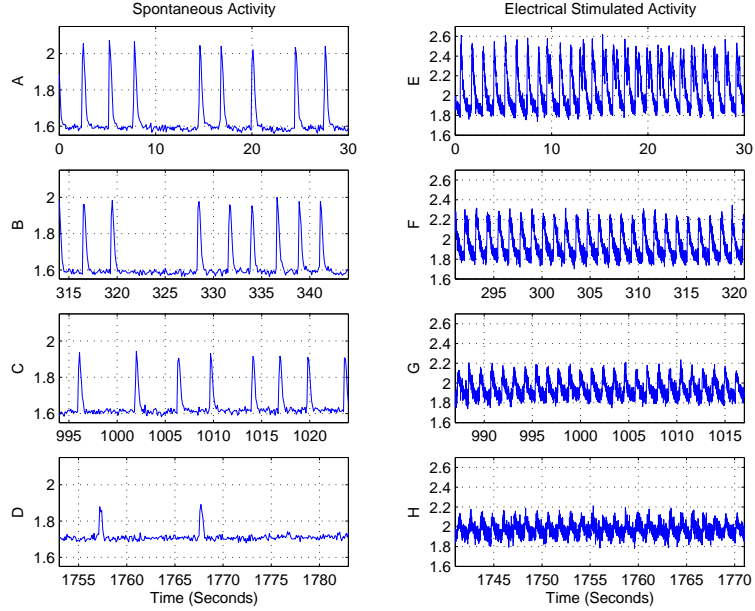,width=10cm}\\
\end{tabular}
\end{center}
\end{minipage}}
\\
\caption{\small Cardiac cells, 4 days in culture, were exposed to
magnetic fields of magnitude 160 pT and frequency 16 Hz for 30 min.
Characteristic traces of  spontaneous cytosolic calcium activity
(A,B,C,D) and of electrically stimulated (1 Hz) cytosolic calcium
activity (E,F,G,H). Times are measure in seconds from the moment of
application of the magnetic field. } \label{f:Asher1}
\end{figure}

The effect of ELF-EMF on cardiac myocytes at the
$I_{\mbox{\scriptsize Ks}}$ was demonstrated in \cite{Asher2007}.
Specifically, neonatal rat cardiac myocytes in cell culture were
exposed to electromagnetic fields at frequencies 15 Hz, 15.5 Hz, 16
Hz, 16.5 Hz and amplitudes of the magnetic field from below 16 pT
and up to 160 nT. In the range 16 pT -- 16 nT both stimulated and
spontaneous activity of the myocytes changed at frequency 16 Hz: the
height and duration of cytosolic calcium transients began decreasing
significantly about 2 minutes after the magnetic field was applied
and kept decreasing for about 30 minutes until it stabilized at
about 30\% of its initial value and its width decreased to
approximately 50\%. About 10 minutes following cessation of the
magnetic field the myocyte (spontaneous) activity recovered with
increased amplitude, duration, and rate of contraction. Outside this
range of frequencies and magnetic fields no change in the transients
was observed (see Figure \ref{f:Asher1}). When the stereospecific
inhibitor of KCNQ1 and $I_{\mbox{\scriptsize Ks}}$ channels
chromanol 293B was applied, the phenomenon disappeared, which
indicates that the $I_{\mbox{\scriptsize Ks}}$ and KCNQ1 potassium
channels in the cardiac myocyte are the targets of the
electromagnetic field, in agreement with the results of the oocyte
experiment mentioned above.

The peaking of the effect of the ELF-EMF at definite frequencies
ties interaction phenomenon to possible stochastic resonance.
Stochastic resonance (SR) constitutes a cooperative phenomenon,
whereby the addition of noise to a periodic signal measured by a
nonlinear bistable (or multistable) system can, paradoxically, boost
the detected signal \cite{Gammaitoni1998}. A voltage-gated ion
channel in a biological membrane can be viewed as a bistable
nonlinear system driven by noise (thermal fluctuations), which
causes random transitions between the open and closed states at
rates that depend on membrane voltage. Thus voltage-gated ionic
channels can be naturally expected to exhibit SR.

Voltage-gated channels are known to gate spontaneously, that is, to
open and close to permeating ions at random times, without the
application of any apparent external force \cite{Hille}. There are
many theories of spontaneous random gating of protein channels of
biological membranes \cite{Sanguinetti1996}, \cite{Yellen2002},
\cite{Horn2000}, \cite{Larsson2002}, \cite{Bezanilla2002},
\cite{Jiang2003}. When an alternating magnetic field is applied
transversally to the channel, an alternating electromotive force is
created, according to Faraday's law, which superimposes an
alternating electric force on the mobile ions in the channel. The
specific response at 16 Hz may indicate some form of resonance or
stochastic resonance of a gating mechanism of open voltage-gated
potassium channels (e.g., a secondary structure or mechanism) with
time-periodic induced electric field. A summary of the SR theory
proposed in \cite{ChemPhys} is given in Section \ref{s:Theory}.

To tie the above considerations to the communicated effects on
cardiac myocytes and on the heart, we note that according to our
theory, the open probability of the relevant K$^+$ channel is
increased during the plateau due to the said resonance. Therefore
the increased efflux of potassium shortens the action potential, and
consequently lowers the peak of the cytosolic calcium concentration,
at the expense of increased sodium concentration (see Figures 9 and
10). The shortening of the action potential leads to the shortening
of the QT interval (see Figure 11) \cite{Opie2003}. These
theoretical predictions are supported by experimental measurements.
Specifically, similar phenomena were communicated in
\cite{Mazhari2002}, \cite{Jeong2005}. Other manifestations of
resonance with magnetic fields at 8 and 16 Hz and amplitudes between
160 pT and 160 nT, consistent with the prediction of the modified
Luo-Rudy model (see figures \ref{f:AP_k127_Vhalf28m_m20_p28_Iks},
\ref{f:APD_k127_Vhalf28m_m20_p28_Iks}),  were communicated in
\cite{Asher2007}. The phenomenon of resonance with an alternating
electric field in a potassium channel was reported first in
\cite{Menconi1998}.

\section{Materials and Methods}

Channel expression into {\em Xenopus} oocytes Female {\em Xenopus
Laevis} frogs were purchased from Xenopus 1 (Dexter, Michigan, USA).
The procedures followed for surgery and maintenance of frogs were
approved by the animal research ethics committee of Tel Aviv
University and in accordance with the Guide for the Care and Use of
Laboratory Animals (1996. National Academy of Sciences, Washington
D.C.). Frogs were anaesthetized with 0.15\% tricaine (Sigma). Pieces
of the ovary were surgically removed and digested with 1mg/ml
collagenase (type IA, Sigma) in Ca$^{2+}$-free ND96 for about one
hour, to remove follicular cells. Stage V and VI oocytes were
selected for cRNA or DNA injection and maintained at 18$^o$C in ND96
(in mM: 96 NaCl, 2 KCl, 1.8 mM CaCl$_2$, 1 MgCl$_2$ and 5 HEPES
titrated to pH = 7.5 with NaOH), supplemented with 1mM pyruvate and
50 µg/ml gentamicin. The human KCNQ1 cDNA (in pGEM vector) was
linearized by Not1. Capped complementary RNA was transcribed by the
T7 RNA polymerase, using the mMessage mMachine transcription kit
(Ambion Corp). The cRNA size and integrity was confirmed by
formaldehyde-agarose gel electrophoresis. Homomeric expression of
human KCNQ1 or I$_{\mbox{\scriptsize Ks}}$ was performed by
injecting 40 nl per oocyte (5 ng cRNA) using a Nanoject injector
(Drummond, USA). Several expression experiments were also carried
out by microinjecting a recombinant DNA vector (pcDNA3) encoding the
human KCNQ1 or I$_{\mbox{\scriptsize Ks}}$ cDNA directly into
Xenopus oocyte nuclei (1 ng into 10 nl). Very similar data were
obtained for either cRNA or DNA injections.

\subsection{Electrophysiology}

Standard two-electrode voltage-clamp measurements were performed at
room temperature (22$^o$C-24$^o$C) 2-5 days following cRNA or DNA
microinjection. Oocytes were placed into a 100 $\mu$l recording
chamber and superfused with a modified ND96 solution (containing
0.1mM CaCl$_2$) under constant perfusion using a fast perfusion
system at a rate of 0.48 ml/sec (ALA VM8, ALA Scientific
Instruments). Whole-cell currents were recorded using a GeneClamp
500 amplifier (Axon Instruments). Stimulation of the preparation,
and data acquisition were performed using the pCLAMP 6.02 software
(Axon Instruments) and a 586 personal computer interfaced with a
Digidata 1200 interface (Axon Instruments). Stimulation of the
preparation, and data acquisition were performed using the pCLAMP
6.02 software (Axon Instruments) and a 586 personal computer
interfaced with a Digidata 1200 interface (Axon Instruments). Glass
microelectrodes (A-M systems, Inc) were filled with 3M  KCl and had
tip resistances of 0.2-1 M$\Omega$. Current signals were digitized
at 1 kHz and low pass filtered at 0.2 kHz. The holding potential was
-80 mV. Leak subtraction was performed off-line, using steps from
-120 to -90 mV, assuming that the channels are closed at -80 mV and
below. Errors introduced by the series resistance of the oocytes
were not corrected and were minimized by keeping expression of the
currents below 10 $\mu$A.

\subsection{Application of the magnetic field}

A digital waveform Agilent generator was connected to a copper
wire coil with a single wrapping of diameter 5 cm and alternating
sinusoidal currents of 16 Hz and amplitude 0.636 mA were passed.
The amplitudes of the generated magnetic fields at the center of
the loop were 16nT, according to the Biot-Savart law. The copper
wire was wound around the oocytes . We repeated the experiments 5
times on separate oocytes.

\subsection{Data analysis}

Data analysis was performed using the Clampfit program (pCLAMP 8,
Axon Instruments), Microsoft Excel 2002 (Microsoft Corporation),
SigmaPlot 8.0 (SPSS Inc) and Prism (GraphPad software). To analyze
the voltage dependence of channel activation, a double exponential
fit was applied to the tail currents at -60 mV or -120 mV and the
slow exponential component was extrapolated to the beginning of the
repolarizing step. Chord conductance (G) was calculated by using the
equation
 $$ G = \frac{I}{V-V_{\mbox{\scriptsize rev}}},$$
where $I$ is the extrapolated tail current and $V_{\mbox{\scriptsize
rev}}$ is the reversal potential measured in each experiment. The
reversal potential value was $V_{\mbox{\scriptsize rev}}=-98 \pm 2
mV$ (n = 10). The conductance $G$ was estimated at the tail voltage
$V$ and then normalized to the maximal conductance value
$G_{\mbox{\scriptsize max}}$. Activation curves were fit to the
Boltzmann distribution
 $$\frac{G}{G_{\mbox{\scriptsize
 max}}} =\frac{1}{1+\exp\left\{\ds\frac{V_{50}-V}{s}\right\}},$$
where $V_{50}$ is the voltage at which the open probability is 50\%,
and $s$ is a slope factor.

\section{Results}
\begin{figure}
\centering
{\includegraphics[width=8cm]{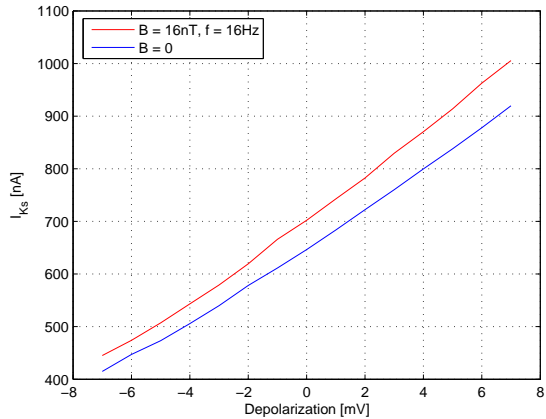}}
\caption{\small $I_{\mbox{\scriptsize Ks}}$ current in {\em Xenopus}
oocytes with applied magnetic field of 16 Hz and 16 nT (red) and
without (blue)} \label{f:redblue}
\end{figure}

\begin{figure}
\mbox{
\begin{minipage} {\textwidth}
\begin{center}
\begin{tabular}{c}
\psfig {figure=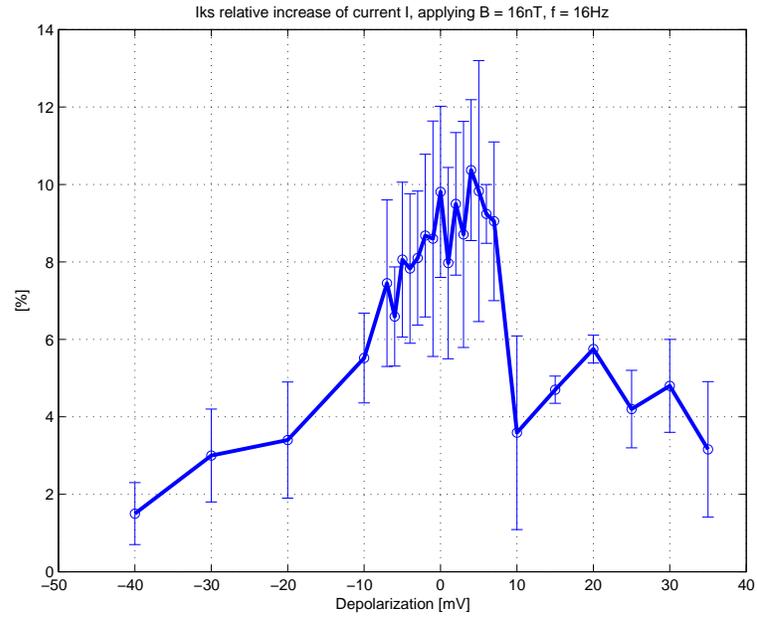,width=10cm}\\
\end{tabular}
\end{center}
\end{minipage}}
\\
\caption{\small Resonant increase in the $I_{\mbox{\scriptsize
Ks}}$ current at magnetic field of 16 Hz and 16 nT}
\label{f:SIKUM_Iks_change_B_16nT_f16_sofi}
\end{figure}
\begin{figure}
\mbox{
\begin{minipage} {\textwidth}
\begin{center}
\begin{tabular}{c}
\psfig {figure=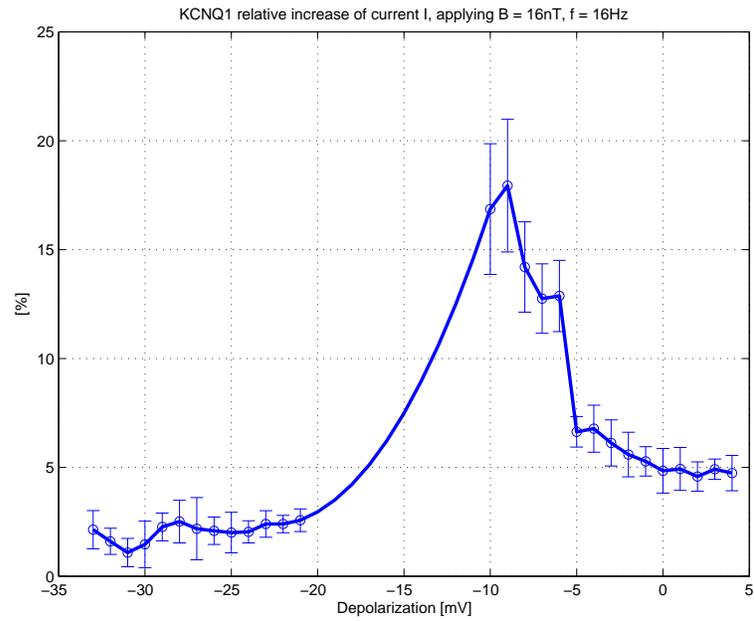,width=10cm}\\
\end{tabular}
\end{center}
\end{minipage}}
\\
\caption{\small Resonant increase in the KCNQ1 channel current at
magnetic field of 16 Hz and 16 nT}
\label{f:SIKUM_KCNQ1_change_B_16nT_f16_sofi}
\end{figure}

A direct measurement of resonance between an applied periodic
magnetic field and the potassium current in a human
$I_{\mbox{\scriptsize Ks}}$ channel, expressed in {\em Xenopus}
oocytes, shows a maximal increase of about 9\% in the current at
frequency 16 Hz and amplitude 16 nT of the applied magnetic field
and membrane depolarization between 0 mV and 8 mV (see Figures
\ref{f:redblue} and \ref{f:AP_k127_Vhalf28m_m20_p28_Iks}). A similar
measurement of the potassium current in the KCNQ1 channel, expressed
in an oocyte, gives a maximal increase of 16\% at the same applied
magnetic field and membrane depolarization between -14 mV and -7 mV
(see Figure \ref{f:SIKUM_KCNQ1_change_B_16nT_f16_sofi}).

\section{Theory}\label{s:Theory}

\begin{figure}
\mbox{
\begin{minipage} {\textwidth}
\begin{center}
\begin{tabular}{c}
\epsfig {figure=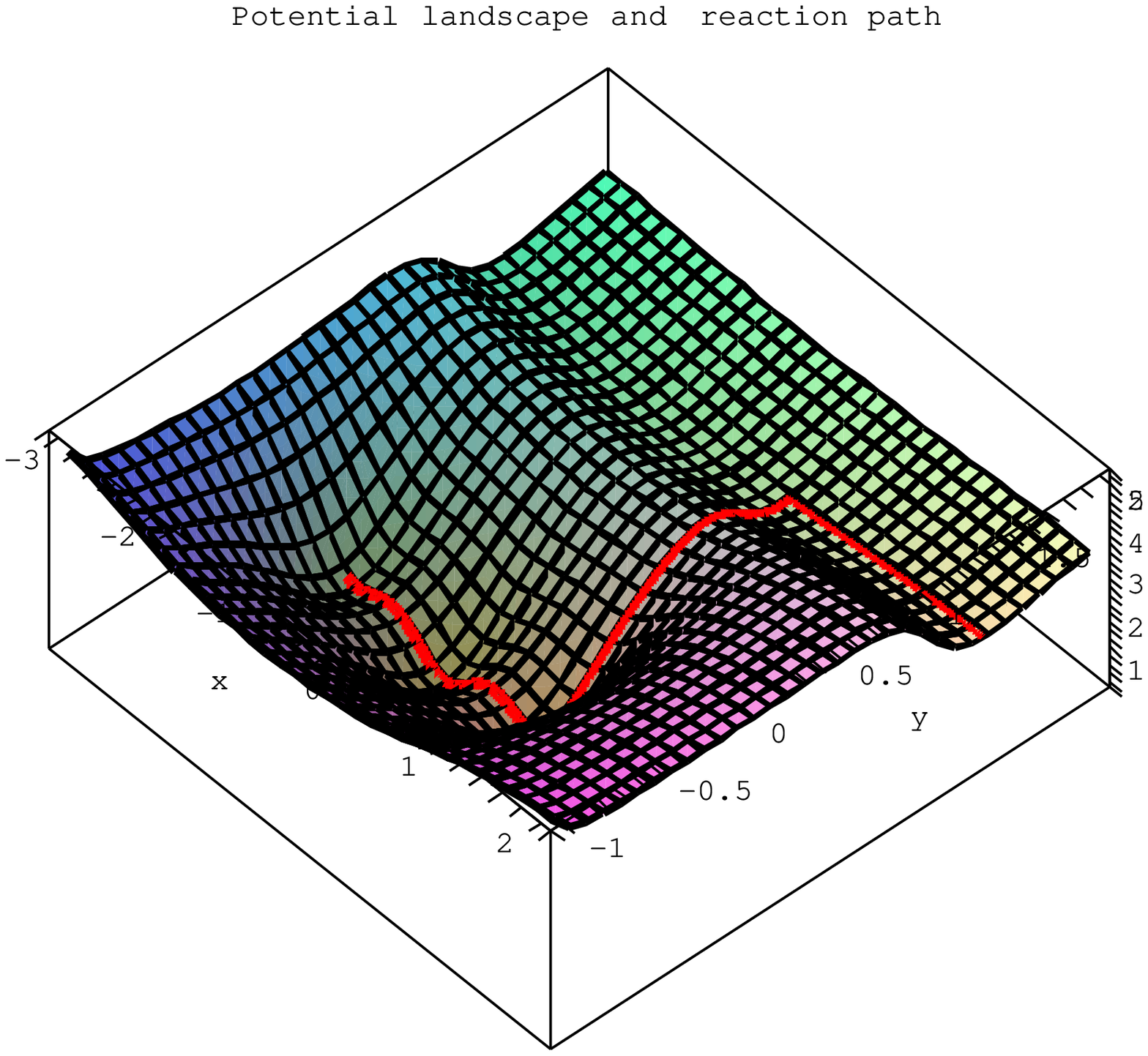,width=8cm}\\
\end{tabular}
\end{center}
\end{minipage}}
\\
\caption{\small Hypothetical energy landscape of two ions in the
selectivity filter. The reaction path is marked red. The straight
segment in the trough may represent the open state in the channel}
 \label{f:potential-path1}
\end{figure}
\begin{figure}
\mbox{
\begin{minipage} {\textwidth}
\begin{center}
\begin{tabular}{c}
\psfig {figure=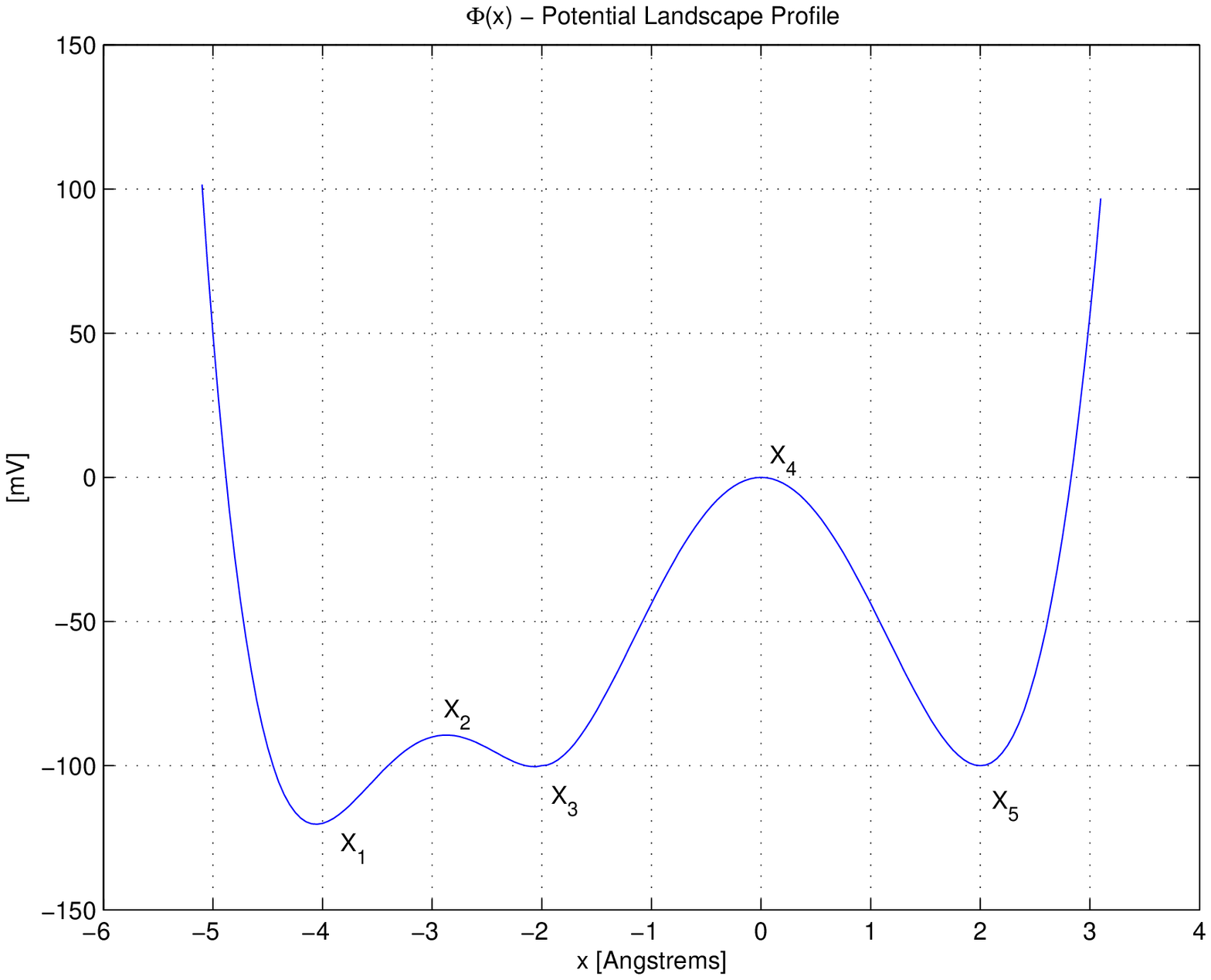,width=7cm}\\
\end{tabular}
\end{center}
\end{minipage}}
\\
\caption{\small Profile of one-dimensional electrostatic potential
landscape biased by a constant electric field}
\label{f:doublewell-landscape}
\end{figure}

Since the induced electric field is too low to interact with any
component of the $I_{\mbox{\scriptsize Ks}}$ channel, we conjecture
that the induced field may interact with locally stable (metastable)
configurations of ions inside the selectivity filter  \cite{Roux}.
We propose in \cite{ChemPhys} an underlying scenario for this type
of interaction based on the collective motion of three ions in the
channel, as represented in the molecular dynamics simulation of
\cite{Roux}. The configurations of three potassium ions in the KcsA
channel is represented in \cite{Roux} in reduced reaction
coordinates on a three-dimensional free energy landscape. In our
simplified model, we represent the collective motion of the three
ions in the channel as diffusion of a higher-dimensional Brownian
particle in configuration space. An imitation hypothetical energy
landscape with a reaction path (indicated in red) is shown in Figure
\ref{f:potential-path1}. Projection onto a reaction path reduces
this representation to Brownian motion on one-dimensional landscape
of potential barriers (see Figure \ref{f:doublewell-landscape}). The
stable states represent instantaneous crystallization of the ions
into a metastable configuration, in which no current flows through
the channel, that is, they represent closed states of the channel.
There is also a pathway in the multidimensional energy landscape
that corresponds to a steady current flowing in the channel, e.g.,
an unobstructed trough in the energy landscape. Transitions from the
latter into the former represent gating events. In the scenario of
\cite{ChemPhys} the motion between closed states is simplified to
one-dimensional Brownian motion, e.g., in a trough obstructed  with
barriers, while the interruptions in the current correspond to exits
from the unobstructed trough into the obstructed one. Activated
transitions over barriers separating two closed states in the
obstructed trough affect the probability of transition from closed
to open states. Stochastic resonance between two closed states may
change the transition rates between them, thus affecting the open
(or closed) probability of the channel. The theoretical prediction
of [34] claim a relatively narrow window of frequencies and
depolarizations, in which the probability of staying on one side of
the barrier peaks to about 10-15\%, depending on amplitude, above
that without the application of the electromagnetic field. These
predictions seem to be consistent with the experimental result
communicated in this paper.

Both the theoretical and experimental results suggest that the
Hodgkin-Huxley equations for channel conductance-voltage response
should be modified in the presence of the above mentioned resonance.
At the resonance frequency of 16 Hz a bump of about 10-16\% should
appear in the conductance-voltage response curve at resonance
membrane depolarizations, as shown in Figure \ref{f:redblue}.
Alternatively, the response curve can be shifted to the left. Some
of the effects of the ELF-EMF on cardiac myocytes, as observed both
experimentally \cite{Asher2007} and predicted by the Luo-Rudy model
\cite{Zeng-Rudy1995} with the modified Hodgkin-Huxley equations, are
shown in Figures 7 and 8. They consist, among others, in the
shortening the duration and lowering the amplitude of the action
potential and of the calcium transients.
\begin{figure}
\mbox{
\begin{minipage} {\textwidth}
\begin{center}
\begin{tabular}{c}
\epsfig {figure=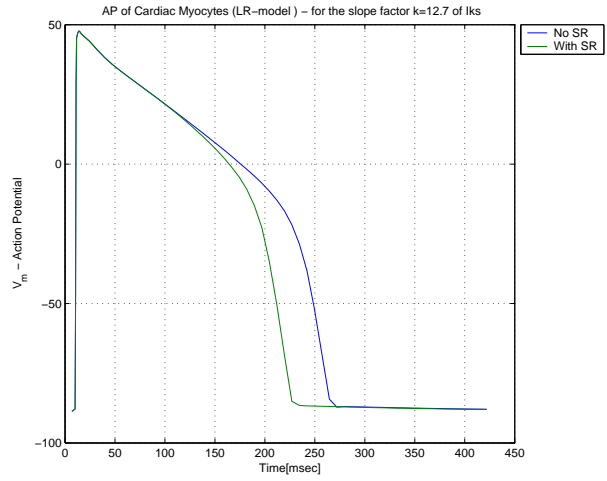,width=8cm}\\
\end{tabular}
\end{center}
\end{minipage}}
\\
\caption{\small Action potential without magnetic field (Blue) and
with magnetic field (Green)} \label{f:AP_k127_Vhalf28m_m20_p28_Iks}
\end{figure}

\begin{figure}
\mbox{
\begin{minipage} {\textwidth}
\begin{center}
\begin{tabular}{c}
\epsfig {figure=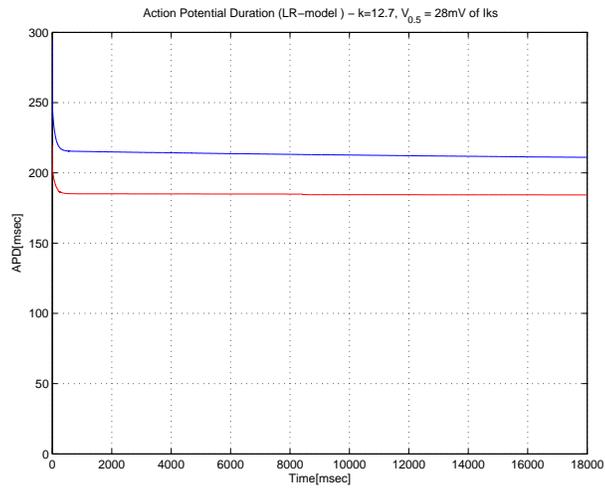,width=8cm}\\
\end{tabular}
\end{center}
\end{minipage}}
\\
\caption{\small Action potential duration without magnetic field
(Blue) and with magnetic field (Red)}
\label{f:APD_k127_Vhalf28m_m20_p28_Iks}
\end{figure}

\section{Conclusion and Discussion}
Following observations of interactions between ELF-EMF in the heart
and in calcium transients in cardiac myocytes, we set out to
localize the effect at the molecular level. We find that the
$I_{\mbox{\scriptsize Ks}}$ and KCNQ1 channels respond to ELF-EMF at
the particular frequency of 16 Hz and at membrane depolarization
range that corresponds to the plateau of the action potential in the
cardiac myocyte. Specifically, we have exposed human
$I_{\mbox{\scriptsize Ks}}$ and KCNQ1 channels, expressed in a {\em
Xenopus} oocyte, to a periodic magnetic field of frequency 16 Hz and
amplitude 16 nT and varied the membrane depolarization between -100
mV and +100 mV. The observed response was a maximal increase of
about 9\% in the potassium current at membrane depolarization
between 0 mV and 8 mV for the $I_{\mbox{\scriptsize Ks}}$ channel
(see Figure \ref{f:SIKUM_Iks_change_B_16nT_f16_sofi}) of 16\% at the
same applied magnetic field and membrane depolarization between -14
mV and -7 mV for the KCNQ1 channel(see Figure
\ref{f:SIKUM_KCNQ1_change_B_16nT_f16_sofi}). To explain this
phenomenon, we offer a scenario of a new kind of stochastic
resonance between the induced periodic electric field and the
thermally activated transitions between locally stable
configurations of the mobile ions in the selectivity filter. Since
the induced electric field is too weak to interact with any
component of the $I_{\mbox{\scriptsize Ks}}$ or KCNQ1 channel
protein, our theory cannot describe  the primary gating mechanism of
a voltage gated channel. We therefore resort to a new scenario,
which postulates interaction of the induced field with
configurations of the mobile ions inside the selectivity filter.
These configurations may be much more susceptible to the weak
induced fields than any components of the surrounding protein,
because the potential barriers separating the metastable
configurations of the mobile ions can be of any height.


\begin{thebibliography}{99}
\bibitem{DiCarlo}A.L. DiCarlo, J. M. Farrell, T. A. Litovitz,
"Myocardial protection conferred by electromagnetic fields", {\em
Circulation} {\bf99}, pp.813--816 (1999).

\bibitem{Sait}M.L. Sait, A.W. Wood,  and H.A. Sadafi, "A study of heart
rate and heart rate variability in human subjects exposed to
occupational levels of 50 Hz circularly polarized magnetic fields",
{\em Med. Eng. Phys.} {\bf21}, pp.361--369 (1999).

\bibitem{Fadel}M.A. Fadel, S.M. Wael, R.M. Mostafa, "Effect of 50 Hz,
0.2 mT magnetic fields on RBC properties and heart functions of
albino rats", {\em Bioelectromagnetics} {\bf24}, pp.535--545 (2003).

\bibitem{Jeong2005} J.H. Jeong, J.S. Kim, B.C. Lee, Y.S. Min, D.S. Kim,
J.S. Ryu, K.S. Soh, K.M. Seo, U.D. Sohn, "Influence of exposure to
electromagnetic field on the cardiovascular system", {\em Autonomic
\& Autacoid Pharmacology} {\bf25} (1), pp.17--23 (7) (2005).

\bibitem{Asher2007}M. Shaked, T. Zinman, A. Shainberg, S. Barzilai,
M. Scheinowitz, S. Laniado, T. Kamil, Z. Schuss, ``The effect of
extremely low frequency and amplitude magnetic fields on the cycle
of cardiac myocytes'', (preprint 2007).

\bibitem{Savitz}D.A. Savitz, D. Liao, A. Sastre, R.C. Kleckner and R.
Kavet, "Magnetic Field Exposure and Cardiovascular Disease Mortality
among Electric Utility Workers", {\em Am. J. of Epidemiol.}
{\bf149}, pp.135--142 (1999).

\bibitem{Korpinen}L. Korpinen  and J. Partanen, "Influence of 50-Hz electric
and magnetic fields on human blood pressure", {\em Radiat Environ.
Biophys.} {\bf35}, pp.199--204 (1996).

\bibitem{Tepper}O.M. Tepper, M.J. Callaghan, E.I. Chang, R.D. Galiano, K.A.
Bhatt, S. Baharestani, "Electromagnetic fields increase in vitro and
in vivo angiogenesis through endothelial release of FGF-2" {\em
FASEB J.} {\bf18}, pp.1231--1233 (2004).

\bibitem{Blackman} Blackman C.F., S.G. Benane, L. Kinney, W. Joines, and D. E.
House, "Effects of ELF fields on calcium-ion efflux from brain
tissue in vitro", {\em Radiat. Res.} {\bf92}, pp.510-520 (1982).

\bibitem{Blackman85}C.F. Blackman, S.G. Benane, J. Rabinowitz, D.E. House, and W.
Joines, "A role of the magnetic field in the radiation induced
efflux of calcium ions from brain tissue in vitro", {\em
Bioelectromagnetics} {\bf6}, pp.327--337 (1985).

\bibitem{Galvanovskis}J. Galvanovskis, J. Sandblom, B. Bergqvist, S. Galt, Y.
Hamnerius, "Cytoplasmic oscillations in human leukemia T-cells are
reduced by 50 Hz magnetic fields", {\em Bioelectromagnetics}
{\bf20}, pp.269--276 (1999).

\bibitem{Rosenspire}A.J. Rosenspire, A.L. Kindzelskii, B.J. Simon,
H.R. Petty, "Real-Time Control of Neutrophil Metabolism by Very Weak
Ultra-Low Frequency Pulsed Magnetic Fields, {\em Biophys. J.}
{\bf88}, pp.3334--3347 (2005).

\bibitem{Baureus}C.L.M. Baur\'eus Koch, M.  Sommarin, B.R.R. Persson, L.G. Salford,
and J.L. Eberhardt, "Interaction between weak low frequency magnetic
fields and cell membranes", {\em Bioelectromagnetics} {\bf24} (6),
pp.395--402. (2003).


\bibitem{McLeod}B.R. McLeod, A.R. Liboff, "Dynamic characteristics of
membrane ions in multifield configurations of low-frequency
electromagnetic radiation", {\em Bioelectromagnetics} {\bf7},
pp.177--189 (1986).

\bibitem{Giudice}E. Del Giudice, M. Fleischmann, G. Preparata, and G.
Talpo, "On the unreasonable effects of ELF magnetic fields upon a
system of ions", {\em Bioelectromagnetics} {\bf23}, pp.522--530
(2002).

\bibitem{Liboff}A.R. Liboff, "Electric-Field Ion Cyclotron Resonance,
{\em Bioelectromagnetics} {\bf18}, pp.85--87 (1997).

\bibitem{Zhadin}M.N. Zhadin, F. Barnes, "Frequency and amplitude windows in
the combined action of DC and low frequency AC magnetic fields on
ion thermal motion in a macromolecule: Theoretical analysis", {\em
Bioelectromagnetics} {\bf26}, pp.323--330 (2005).

\bibitem{Adair}R.K. Adair, "Comment: Analyses of models of ion actions
under the combined action of AC and DC magnetic fields", {\em
Bioelectromagnetics} {\em27}, pp.332--334 (2006).

\bibitem{Sanguinetti1991} M. C. Sanguinetti and N. K. Jurkiewicz,
"Delayed rectifier outward K$^+$ current is composed of two currents
in guinea pig atrial cells", {\em Am. J. Physiol. Heart Circ.
Physiol.} {\bf260}: H393--H399 (1991) 0363--6135/91.

\bibitem{Barhanin1996}J. Barhanin, F. Lesage, E. Guillemare, M. Fink,
M. Lazdunski, and G. Romey, ``K(v)LQT1 and $I_{\mbox{\scriptsize
Ks}}$ (minK) proteins associate to form the $I_{\mbox{\scriptsize
Ks}}$ cardiac potassium current'', {\em Nature}, {\bf384}, pp.78–80
(1996).

\bibitem{Barhanin2000}N. Tinel, S. Diochot, M. Borsotto, M. Lazdunski, and
J. Barhanin, ``KCNE2 confers background current characteristics to
the cardiac KCNQ1 potassium channel'', {\em EMBO J.} {\bf19} (23),
pp.6326–-6330 (2000).

\bibitem{Jentsch2000}B.C. Schroeder, S. Waldegger, S. Fehr, M. Bleich, R. Warth,
R. Greger, and T.J. Jentsch,  ``A constitutively open potassium
channel formed by KCNQ1 and KCNE3'', {\em Nature} {\bf403},
pp.196-–199 (2000).

\bibitem{Nerbonne2005}J.M. Nerbonne and R.S. Kass, ``Molecular Physiology
of Cardiac Repolarization'', {\em Physiol. Rev.} {\bf85},
pp.1205--1253 (2005).

\bibitem{Mazhari2002} R. Mazhari, H.B. Nuss, A.A. Armoundas, R.L. Winslow,
E. Marban, "Ectopic expression of KCNE3 accelerates cardiac
repolarization and abbreviates the QT interval", {\em J. Clin.
Invest.} {\bf109}, pp.1083--1090 (2002).

\bibitem{Roden2006}D.M. Roden, ``A New Role for Calmodulin in Ion Channel Biology'',
{\em Circ. Res.} {\bf98}, pp.979 (2006).

\bibitem{Gammaitoni1998} L. Gammaitoni, P. H\"anggi, P. Jung, and F.
Marchesoni, "Stochastic Resonance", {\em Rev. Mod. Phys.} {\bf70},
pp.223--288 (1998).

\bibitem{Hille} B. Hille, {\em Ionic Channels of Excitable Membranes} (3rd
edition), Sinauer Assoc. Inc, Sunderland, Mass. 2001.

\bibitem{Sanguinetti1996} M.C. Sanguinetti, M.E. Curran, A. Zou, J. Shen,
P.S. Spector, D.L. Atkinson, and M.T. Keating, ``Coassembly of
KvLQT1 and minK (IsK) to form cardiac $I_{\mbox{\scriptsize Ks}}$
potassium channel'', {\em Nature} {\bf 384}, pp.80--83 (1996).


\bibitem{Yellen2002} G. Yellen, "The voltage--gated potassium channels and their
relatives", {\em Nature} {\bf419}, pp.35--42  (2002).

\bibitem{Horn2000} R. Horn, "A New Twist in the Saga of Charge Movement in
Voltage--Dependent Ion Channels", {\em Neuron} {\bf25}, pp.511--514
(2000).

\bibitem{Larsson2002} H.P. Larsson, "The Search Is on for the Voltage
Sensor--to--gate Coupling", {\em J. Gen. Physiol.} {\bf120},
pp.475--481 (2002).

\bibitem{Bezanilla2002} F. Bezanilla, "Voltage sensor movements", {\em J. Gen.
Physiol.} {\bf120}, pp.465--473 (2002).

\bibitem{Jiang2003} Y. Jiang, A. Lee, J. Chen, V. Ruta, M. Cadene, B.T.
Chait, and R. MacKinnon, "X--ray structure of a voltage--dependent
K1 channel", {\em Nature} {\bf423}, pp.33--41 (2003).

\bibitem{ChemPhys}M. Shaked and Z. Schuss, "Stochastic resonance with applied and
induced fields: the case of voltage-gated ion channels",
http://arxiv.org/abs/0903.0506v1 {\em Chem. Phys.} (submitted 2009).

\bibitem{Opie2003} L.H. Opie, {\em Heart Physiology: From Cell to Circulation},
Lippincott, Williams \& Wilkins; 4th edition 2003.

\bibitem{Menconi1998} M.C. Menconi. M. Pellegrini, M. Pellegrino, D.
Petracchi, "Periodic forcing of a single ion channel: dynamic
aspects of the open--closed switching", {\em Eur. Biophys. J.}
{\bf27}, pp.299--304 (1998).

\bibitem{Roux}S. Bern\`eche and B. Roux, "Energetics of ion conduction
through the K$^+$ channel", {\em Nature}, {\bf414}, pp.73-77 (2001).

\bibitem{Zeng-Rudy1995}J. Zeng, K.R. Laurita, D.S. Rosenbaum, Y. Rudy,
``Two components of the delayed rectifier K$^+$ current in
ventricular myocytes of the Guinea pig type", {\em Circ. Res.}
{\bf77}, pp.140--152 (1995).

%
%
%
%
%
%
%
%
%
%
%
%
%
%
%
%
%
%
%
%
%
%
%
%
%
%
%
%
%
%
%
%
%
%
%
%
%
%
%
%
%
%
%
%
%
%
%
%
%
%
%
%
%
%
%
%
%
%
%
%
%
%
%
%
\end{thebibliography}
\end{document}